\documentclass[twocolumn,journal]{IEEEtran}
\usepackage{longtable}
\usepackage{color}
\usepackage{amsmath}
\usepackage{cancel}
\usepackage{amssymb}
\usepackage{optidef}
\usepackage{lscape}
\usepackage{graphicx}
\usepackage{subcaption}
\usepackage{longtable}
\usepackage{multirow}
\usepackage{accents} 
\usepackage{algorithm,algpseudocode}
\usepackage{dblfloatfix}
\usepackage{array}
\usepackage[labelsep=period]{caption}
\usepackage{setspace}
\usepackage{float}
\usepackage{subfig}
\usepackage{lettrine}
\usepackage[noadjust]{cite}
\usepackage{balance}
\usepackage[english]{babel}
\usepackage[letterpaper, left=0.625in, right=0.625in, bottom=1in, top= 0.75in]{geometry}
\usepackage[figurename=Fig.]{caption}
\usepackage{suffix}
\usepackage{mathtools}
\usepackage{comment} 

\usepackage{mathabx}
\usepackage[dvipsnames]{xcolor}
\renewcommand{\figurename}{Fig.}

\usepackage{geometry} 
\usepackage{amsmath}
\usepackage{xparse}

\ExplSyntaxOn
\NewDocumentCommand{\MeijerG}{smmmm}
{
	\IfBooleanTF{#1}
	{
		\vic_meijerg:nnnnnn { #2 } { #3 } { #4 } { #5 } { small } { }
	}
	{
		\vic_meijerg:nnnnnn { #2 } { #3 } { #4 } { #5 } { } { \; }
	}
}

\seq_new:N \l__vic_meijerg_args_in_seq
\seq_new:N \l__vic_meijerg_args_out_seq

\cs_new_protected:Nn \vic_meijerg:nnnnnn
{
	\seq_set_split:Nnn \l__vic_meijerg_args_in_seq { | } { #3 }
	\seq_clear:N \l__vic_meijerg_args_out_seq  
	\seq_map_inline:Nn \l__vic_meijerg_args_in_seq
	{
		\seq_put_right:Nn \l__vic_meijerg_args_out_seq
		{
			\begin{#5matrix} ##1 \end{#5matrix}
		}
	}
	G\sp{#1}\sb{#2}
	\left(
	\seq_use:Nn \l__vic_meijerg_args_out_seq { #6\middownlike|#6 }
	#6\middownlike|#6
	#4
	\right)
}
\ExplSyntaxOff
\def\BibTeX{{\text B\kern-.05em{\sc i\kern-.025em b}\kern-.08em
		T\kern-.1667em\lower.7ex\hbox{E}\kern-.125emX}}
\usepackage{cleveref}
\definecolor{maroon}{RGB}{186,0,0}
\definecolor{purple}{RGB}{96,26,149}
\definecolor{mavi}{RGB}{46,76,255}
\definecolor{haki}{RGB}{38,99,33}

\usepackage{tikz}
\usepackage{mathdots}
\usepackage{cancel}
\usepackage{color}
\usepackage{siunitx}
\usepackage{array}
\usepackage{multirow}
\usepackage{amssymb}
\usepackage{tabularx}
\usepackage{booktabs}
\usetikzlibrary{fadings}
\usetikzlibrary{patterns}
\usetikzlibrary{shadows.blur}
\usetikzlibrary{shapes}

\begin{document}
 	\title
    {WALoMA: A Multitask Wireless Foundation Model\\via Adaptive Low-Rank Masked Autoencoders}
  \author{Madi Makin,~\IEEEmembership{Graduate Student Member,~IEEE}, 
            Asmaa Abdallah,~\IEEEmembership{Senior Member,~IEEE},\\
            Abdulkadir Celik,~\IEEEmembership{Senior~Member,~IEEE},
            and
            Ahmed M. Eltawil,~\IEEEmembership{Senior~Member,~IEEE} 

  \thanks{M. Makin, A. Abdallah, and A. M. Eltawil are with Computer, Electrical, and Mathematical Sciences \& Engineering (CEMSE) Division at King Abdullah University of Science and Technology (KAUST), Thuwal, KSA 23955-6900.}
  \thanks{A. Celik is with the School of Electronics and Computer Science, University of Southampton, Southampton SO17 1BJ, U.K.}

\vspace{-0.7cm}}\maketitle

\begin{abstract}
This paper proposes a multitask wireless foundation model via adaptive low-rank masked autoencoders (WALoMA), a unified multi-task foundation model for sixth-generation (6G) wireless physical layer architectures, to address the limitations of specialized, task-specific deep learning models and the practical challenge of scarce labeled wireless datasets. By leveraging concepts inspired by foundation models, the proposed framework adopts a masked autoencoder (MAE) paradigm to learn from unlabeled channel data, to significantly reduce reliance on extensive annotations. The model treats wireless channel state information (CSI) as a universal modality and learns transferable representations through self-supervised channel reconstruction. Key architectural novelties include the use of 2D positional encoding (PE) to explicitly preserve the spatial-frequency relationships between antennas and subcarriers, and low-rank adaptation (LoRA) for parameter-efficient fine-tuning. The framework's efficacy is demonstrated across five downstream tasks, achieving individual scores of 96.47\% for LoS/NLoS classification, 80.45\% for beam prediction, 85.78\% for channel interpolation, 99.12\% for channel estimation, and 77.18\% for channel charting. Consequently, numerical results show that the proposed model achieves a composite score of 87.80\%, significantly outperforming the 59.90\% achieved by the large wireless model (LWM) baseline while training an average of only 14.68\% of total parameters, and maintaining strong performance even under extremely limited labeled data conditions.
\end{abstract}



\maketitle
\section{Introduction}

\IEEEPARstart{T}{ransitioning} to sixth-generation (6G) wireless networks represents more than an incremental evolution, but a fundamental architectural transformation driven by unprecedented requirements for ultra-reliable low-latency communications (URLLC), extreme data rates, massive connectivity, and intelligent edge services \cite{shamsabadi_6g,prasad_6g}. Achieving these performance targets demands a radical rethinking of physical layer (PHY) design, where conventional model-based signal processing techniques are increasingly strained by the scale, heterogeneity, and dynamic nature of emerging wireless environments.

Modern PHY architectures heavily rely on massive multiple-input multiple-output (MIMO), millimeter-wave and terahertz communications, and orthogonal frequency division multiplexing (OFDM). While these technologies unlock significant spectral efficiency gains, they also introduce high-dimensional optimization problems and intricate channel dynamics \cite{chen_compl,gkonis_compl}. Core tasks such as channel estimation, beamforming design, signal detection, and resource allocation become computationally intensive and analytically intractable under realistic propagation conditions, hardware impairments, and mobility scenarios \cite{gkonis_compl, islam_compl}. As a result, traditional closed-form or iterative optimization methods face scalability and adaptability limitations.

Data-driven deep learning (DL) has therefore emerged as a powerful alternative, capable of modeling complex non-linear relationships directly from high-dimensional wireless data \cite{abdallah_dl1,abdallah_dl2}. Despite promising performance gains across individual PHY-layer tasks, most existing DL-based solutions remain narrowly specialized. Models are typically designed for a single objective and trained under fixed system configurations, requiring retraining when antenna geometries, channel statistics, or deployment scenarios change \cite{vahid_dl}. Such task-specific designs limit generalization, increase deployment overhead, and hinder scalability in dynamic 6G environments.

Addressing these limitations requires unified and adaptable learning frameworks capable of jointly handling multiple interrelated wireless tasks within a single architecture. Multitask large language models (LLMs) and foundation models offer a promising direction by learning shared representations across heterogeneous communications and signal processing objectives \cite{liu_llm4wm,yang_multillm,zheng_multillm}. By enabling cross-task knowledge transfer, such models can improve generalization, reduce retraining costs, and support more scalable and intelligent PHY-layer design for future 6G systems. Inspired by the success of LLMs and vision transformers (ViTs) \cite{dosovitskiy_vit}, these frameworks treat wireless channel state information (CSI) as a universal modality, analogous to text tokens or image patches. Through large-scale pretraining, they aim to learn transferable representations that can be efficiently adapted to diverse downstream tasks with minimal additional training.

Despite these advances, several critical challenges remain. The scarcity of labeled wireless datasets restricts the scalability of supervised learning approaches. Moreover, existing models are inherently task-specific and fail to generalize across heterogeneous deployment scenarios that necessitate repeated retraining and increasing system complexity. Finally, the lack of a unified framework capable of learning transferable, task-agnostic representations of wireless channels limits the realization of truly intelligent PHY-layer architectures. These challenges call for a data-efficient and scalable learning paradigm that can exploit the intrinsic structure of wireless signals while minimizing reliance on labeled data, which is the main source of motivation for this paper.

\subsection{Related Works}
The first wave of DL integration focused on swapping out individual analytical blocks for neural networks. Convolutional neural networks (CNNs) became a building block for CSI compression and feedback \cite{cao_cnn2,guo_cnn1,fan_cnn3}, due to their capability of treating time-frequency channel responses as images. On the other side, recurrent neural networks (RNNs) were similarly drafted for channel prediction \cite{zhu_rnn1,lemayian_rnn2,jiang_rnn3}. Because standard RNNs struggle with long-term dependencies \cite{ashish_attention}, long short-term memory (LSTM) networks were widely adopted to mitigate vanishing gradients and better retain historical channel states \cite{luo_lstm, helmy_lstm, nguyen_lstm}. However, the sequential processing nature of both RNNs and LSTMs inherently bottlenecks strict real-time latency and limits the simultaneous modeling of global spatial-frequency features \cite{ashish_attention}. DL has also been applied to resource allocation, including multi-agent reinforcement learning for reconfigurable intelligent surface (RIS) codebook design \cite{abdallah2024multi,abdallah2024multi1} and power control \cite{fabiani_dlpc}. In this context, while classical optimization and non-DL heuristics provide high reliability and theoretical guarantees for power allocation as explored in \cite{makin1, makin_ojcoms}, DL approaches seek to approximate these solutions with lower online latency. Despite these advances, traditional DL models remain rigid, typically designed for fixed system dimensions and requiring architectural redesign when network configurations change.

Following the paradigm shift in natural language processing (NLP) \cite{ashish_attention} and computer vision (CV) \cite{dosovitskiy_vit}, the wireless community has begun exploring transformer architectures that utilize self-attention to capture long-range dependencies in spatial and frequency domains. Rather than relying on fixed receptive fields, transformers offer a flexible architecture for modeling complex CSI distributions. Driven by the self-attention mechanism, they capture global structural patterns by allowing each patch of the channel matrix to attend to all other visible patches, resolving the long-range dependency issues of CNNs and LSTMs. Expanding on the generative capabilities of physical layer reconstruction, in \cite{liu_found}, a foundation model pretrained via masked channel modeling (MCM) demonstrated that it can effectively extract generic features applicable to diverse downstream tasks that range from channel estimation to beam management. This drive toward unified representation learning has catalyzed the development of large-scale wireless foundation models and adapted language architectures. For example, WirelessGPT \cite{yang_multillm} leverages unsupervised pretraining on massive channel datasets to extract universal spatiotemporal features, enabling seamless adaptation to integrated sensing and communication (ISAC) objectives. In parallel, frameworks such as LLM for wireless multi-tasking (LLM4WM) \cite{liu_llm4wm} adapt pre-trained models for channel-associated multi-tasking by employing a mixture of experts with low-rank adaptation (MoE-LoRA) to align complex wireless data with semantic feature spaces. Similarly, agentic frameworks such as ENWAR \cite{enwar1, enwar2} demonstrated the potential of multimodal LLMs to execute situation-aware reasoning for ISAC and beam tracking. This unsupervised reconstruction-based paradigm has also been extended to cross-layer applications; for instance, the authors in \cite{cheraghinia_found} developed a BERT-based foundation model capable of simultaneous wireless technology recognition and localization, proving that a single pretrained backbone can effectively distinguish between diverse signal types while performing precise ranging. To address the heterogeneity of reconstruction targets across distinct deployment scenarios, the work in \cite{zheng_musefm} introduced multi-task environment-aware foundation model (MUSE-FM), which employs a prompt-guided encoder-decoder mechanism. By treating environmental contexts as multi-modal prompts, this architecture allows for the adaptive reconstruction of channel features across varying signal formats and environments, further validating the efficacy of generative pre-training for the physical layer.

The large wireless model (LWM) \cite{lwm_chall} is widely considered a pivotal first step toward foundation models for wireless communications. While LWM represents a more specialized and effective approach for the physical layer than general-purpose pre-trained models such as GPT or Llama, there remains significant potential for further architectural refinements. Advancing wireless foundation models necessitates architectural refinements that accommodate the unique properties of complex-valued channel representations and varying system dimensions. Improved tokenization strategies and scalable input embeddings can enhance adaptability across heterogeneous deployments. The wireless masked autoencoder (WiMAE) \cite{guler_wimae} exemplifies this evolution by integrating masked autoencoding with contrastive objectives. By leveraging heavily masked inputs (up to 90\%) and treating noisy channel realizations as positive pairs, WiMAE promotes invariant feature learning, resulting in superior linear separability and robustness under low-SNR conditions.

While the LWM established a baseline across multiple tasks, its encoder-only design lacks a generative pretraining mechanism, limiting representation learning under constrained supervision. WiMAE, although adopting masked autoencoding, remains limited in scope and does not generalize across diverse tasks, and falls short of a true wireless foundation model. In contrast, this work introduces a unified foundation model framework, driven by a multitask wireless foundation model via adaptive low-rank masked autoencoders (WALoMA) backbone, that goes beyond incremental architectural improvements. We jointly design self-supervised MAE pretraining and geometry-aware representation learning via: (i) 2D positional encoding (PE) and (ii) scalable adaptation through LoRA within a single cohesive pipeline. This integration enables the model to learn transferable representations from unlabeled data, generalize across heterogeneous system configurations, and efficiently adapt to multiple downstream tasks without retraining the full model. Unlike prior approaches that address these components in isolation, the proposed framework unifies them into a data-efficient and scalable architecture evaluated across five PHY-layer tasks, directly addressing the key limitations of generalization, scalability, and labeled data scarcity.

\subsection{Main Contributions}
This paper introduces a unified and data-efficient foundation model framework for wireless channels, referred to as WALoMA, designed to generalize across diverse physical layer tasks. Its effectiveness is demonstrated through a comprehensive evaluation spanning LoS/NLoS classification, beam prediction, channel interpolation, channel estimation, and channel charting. The main contributions are summarized as follows:

\begin{itemize}
    \item We propose a unified foundation model pipeline that transforms raw wireless channel measurements from heterogeneous sources into invariant token representations, enabling an encoder--decoder transformer to generalize across diverse scenarios.
    
    \item We introduce a geometry-aware representation learning scheme based on 2D PE, which preserves the intrinsic spatial--frequency structure of wireless channels and enables seamless adaptation to varying antenna and subcarrier configurations.
    
    \item To enable efficient multi-task adaptation with minimal trainable parameters, we integrate low-rank adaptation (LoRA) as a parameter-efficient fine-tuning strategy by injecting low-rank updates into the transformer attention layers while freezing the pretrained backbone, enabling efficient multi-task adaptation with minimal trainable parameters.

    \item We demonstrate the robust cross-band generalization of the proposed architecture through highly effective transfer learning from sub-6 GHz datasets to millimeter wave (mmWave) deployment scenarios, supported by extensive analysis showing that 2-dimensional positional encoding (2D PE) drives the largest performance gains and unsupervised channel charting preserves physical topology. Consistently outperforming raw-data approaches even in low-data regimes, the model trains only 14.68\% of its total parameters while achieving strong individual performance across the five downstream tasks (96.47\% for LoS/NLoS classification, 80.45\% for beam prediction, 85.78\% for channel interpolation, 99.12\% for channel estimation, and 77.18\% for channel charting.), culminating in a composite score of 87.80\% that significantly surpasses the LWM baseline (59.90\%).
\end{itemize}

\section{System Model and Problem Formulation}
This section introduces the system model and formulates the multi-task learning problem addressed in this work. We consider wireless CSI as a high-dimensional signal exhibiting structured dependencies across spatial (antenna) and spectral (subcarrier) domains. The objective is to learn a unified representation of the wireless channel that generalizes across heterogeneous deployment scenarios and supports multiple downstream tasks.

We consider a massive multiple-input single-output (MISO) OFDM communication system, where a base station (BS) equipped with a uniform linear array (ULA) of $N_{\rm ant}$ antennas serves a single-antenna user equipment (UE). The frequency-domain channel response is represented by a complex-valued matrix $\mathbf{H} \in \mathbb{C}^{N_{\rm ant} \times N_{\rm sub}},$ where $N_{\rm sub}$ denotes the number of subcarriers. In practical deployments, both $N_{\rm ant}$ and $N_{\rm sub}$ may vary across scenarios, leading to heterogeneous channel dimensions.

We consider a family of downstream wireless tasks defined over channel realizations. Let $\mathcal{T} = \{T_1, T_2, \dots, T_K\}$
denote a set of $K$ tasks, such as LoS/NLoS classification, beam prediction, channel interpolation, channel estimation, and channel charting. Each task $T_k$ defines a mapping
$T_k: \mathbb{C}^{N_{\rm ant} \times N_{\rm sub}} \rightarrow\mathcal{Y}_k$, where $\mathcal{Y}_k$ represents the task-specific output space. Our objective is to learn a unified encoder function $f_\theta: \mathbf{H} \mapsto \mathbf{Z}$, that maps variable-sized channel matrices to a fixed-dimensional latent representation $\mathbf{Z}$. This representation should capture the intrinsic spatial and frequency correlations of the wireless channel and generalize across tasks and deployment scenarios. For each downstream task $T_k$, a compact task-specific head $g_\phi^{(k)}(\cdot)$ is attached such that $\hat{\mathbf{y}}_k = g_\phi^{(k)}\big(f_\theta(\mathbf{H})\big)$. Instead of training separate models for each task independently, we aim to pre-train a single, shared backbone $f_\theta(\cdot)$ that learns transferable channel representations in a self-supervised manner. Lower retraining costs and enhanced performance on unseen data are key advantages of this technique, ultimately paving the way for flexible, multi-task scaling.

\section{Solution Methodology}
This section presents the proposed unified learning framework for data-efficient multi-task modeling of wireless channels. The core idea is to leverage self-supervised learning to pre-train a shared encoder that captures the intrinsic spatial--frequency structure of CSI without labeled data. Specifically, we adopt a \textbf{\textit{channel reconstruction}} paradigm, where the model learns to recover corrupted channel realizations, to enable the extraction of rich, task-agnostic latent representations.

\begin{figure}[ht!]
\centering
\includegraphics[width=0.95\columnwidth]{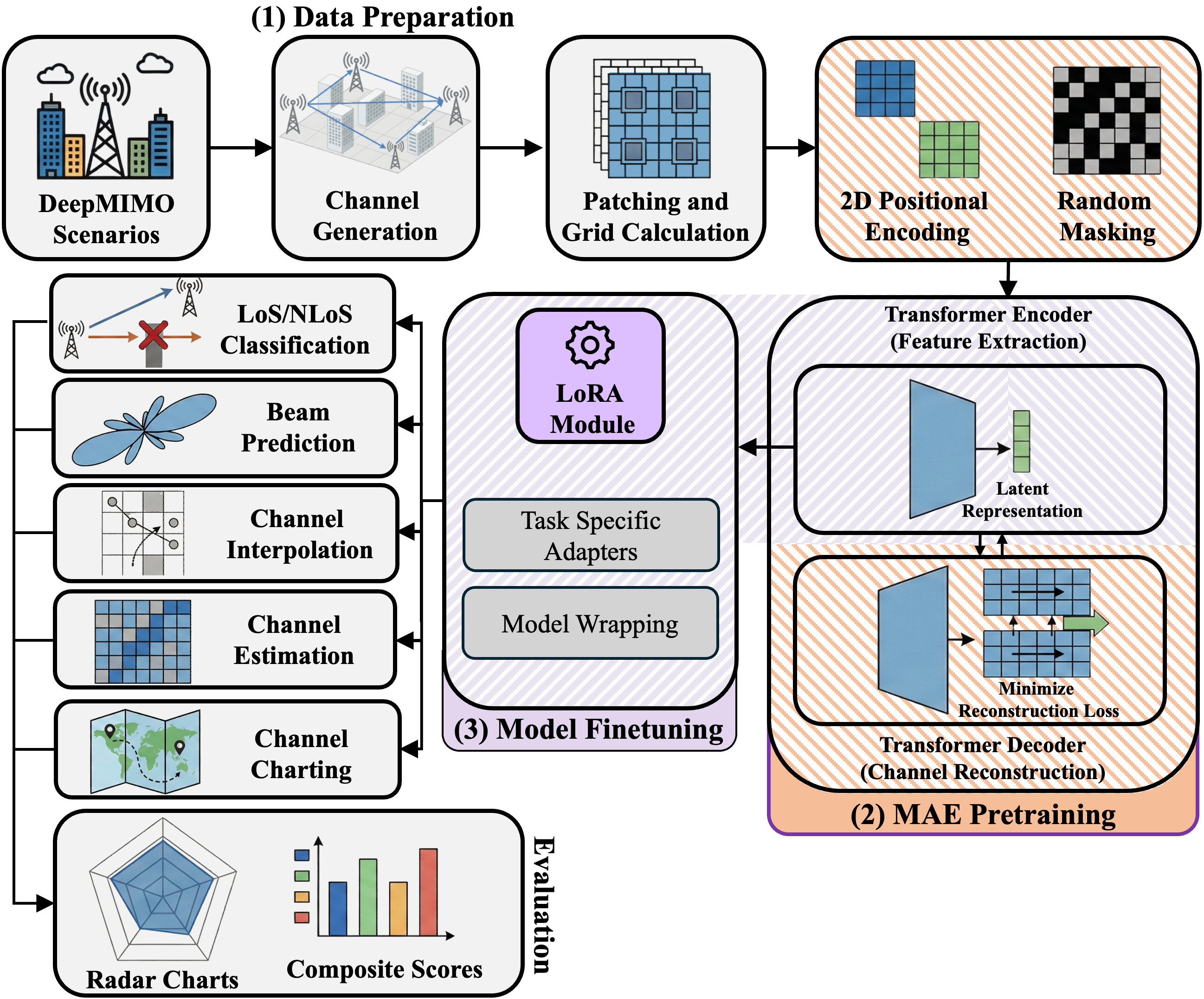}
\caption{Proposed Pipeline.}
\label{fig:sys_mod}
\end{figure} 

\subsection{Self-Supervised Pretraining via Channel Reconstruction}

Wireless channels exhibit structured dependencies across antennas and subcarriers due to spatial correlation, frequency selectivity, and propagation geometry. To capture these intrinsic properties without relying on labeled data, we adopt a masked autoencoder (MAE) pretraining strategy based on channel reconstruction.



Let $\mathbf{M}_m \in \{0,1\}^{N_{\rm ant} \times N_{\rm sub}}$ denote a binary masking matrix generated by a random masking operator that removes a fraction $m$ of channel patches, where ${\mathbf{M}_m}{(i,j)}=0$ indicates a masked entry and ${\mathbf{M}_m}{(i,j)}=1$ denotes a visible entry. The masked input is
\begin{equation}
\tilde{\mathbf{H}} = \mathbf{H} \odot \mathbf{M}_m,
\end{equation}
where $\odot$ denotes element-wise multiplication.

The encoder takes the visible subset $\tilde{\mathbf{H}}$ as input and produces latent embeddings, while a lightweight decoder reconstructs the full channel as
\begin{equation}
\hat{\mathbf{H}} = g_{\psi}\!\left(f_{\theta}(\tilde{\mathbf{H}})\right),
\end{equation}
where $f_{\theta}(\cdot)$ and $g_{\psi}(\cdot)$ denote the encoder and decoder, respectively. The pretraining objective minimizes the reconstruction error over the masked entries only:
\begin{equation}
\mathcal{L}_{\text{pre}}
=
\mathbb{E}_{\mathbf{H},\mathbf{M}_m}
\left[
\left\|
\left(\mathbf{H}-\hat{\mathbf{H}}\right)\odot (1-\mathbf{M}_m)
\right\|_2^2
\right].
\end{equation}
By optimizing reconstruction only over masked regions, the encoder is forced to infer missing channel components from global context. This encourages the model to internalize the spatial–frequency structure of the channel rather than memorize input samples. 

\subsection{Unified Pipeline Overview}

The overview pipeline is presented in Fig. \ref{fig:sys_mod}, where the flow is divided into three main stages: data preparation, MAE pretraining, and model finetuning and evaluation.

\textit{\textbf{Stage 1 - Data Preparation:}} The pipeline begins with raw channel realizations extracted from a given deployment scenario. Each complex-valued channel matrix is decomposed into its real and imaginary components and interpreted as a structured two-dimensional grid over antennas and subcarriers. Rather than flattening the channel into a one-dimensional sequence, the grid is partitioned into small non-overlapping patches. These patches serve as tokens for transformer processing, preserving local spatial–frequency relationships while enabling global context modeling through self-attention.

\textit{\textbf{Stage 2 - MAE Pretraining:}} During pretraining, a 2D positional encoding and high masking ratio is applied to the patch sequence, to preserve the positional knowledge of the tokens and removing a large fraction of tokens from the encoder input. The encoder processes only the visible patches and learns context-aware representations through stacked self-attention layers. A lightweight decoder reconstructs the masked patches from the latent embeddings. This reconstruction-based pretext task compels the encoder to learn robust, scenario-agnostic channel representations that capture global dependencies across antennas and subcarriers.

\textit{\textbf{Stage 3 - Model Finetuning and Evaluation:}} After pretraining, the reconstruction decoder is discarded, and the task-agnostic encoder serves as a shared backbone. To adapt the encoder’s generic understanding of the channels to specific classification and regression objectives, we apply LoRA alongside task-specific heads. Because the encoder has already learned structured channel features through self-supervised reconstruction, this targeted fine-tuning requires significantly fewer labeled samples and converges more rapidly compared to training from scratch. This unified framework enables efficient multi-task adaptation while preserving the underlying physical structure of the wireless environment.

\section{Proposed Design Architecture}

\begin{figure*}[ht!]
\centering
\includegraphics[width=1.7\columnwidth]{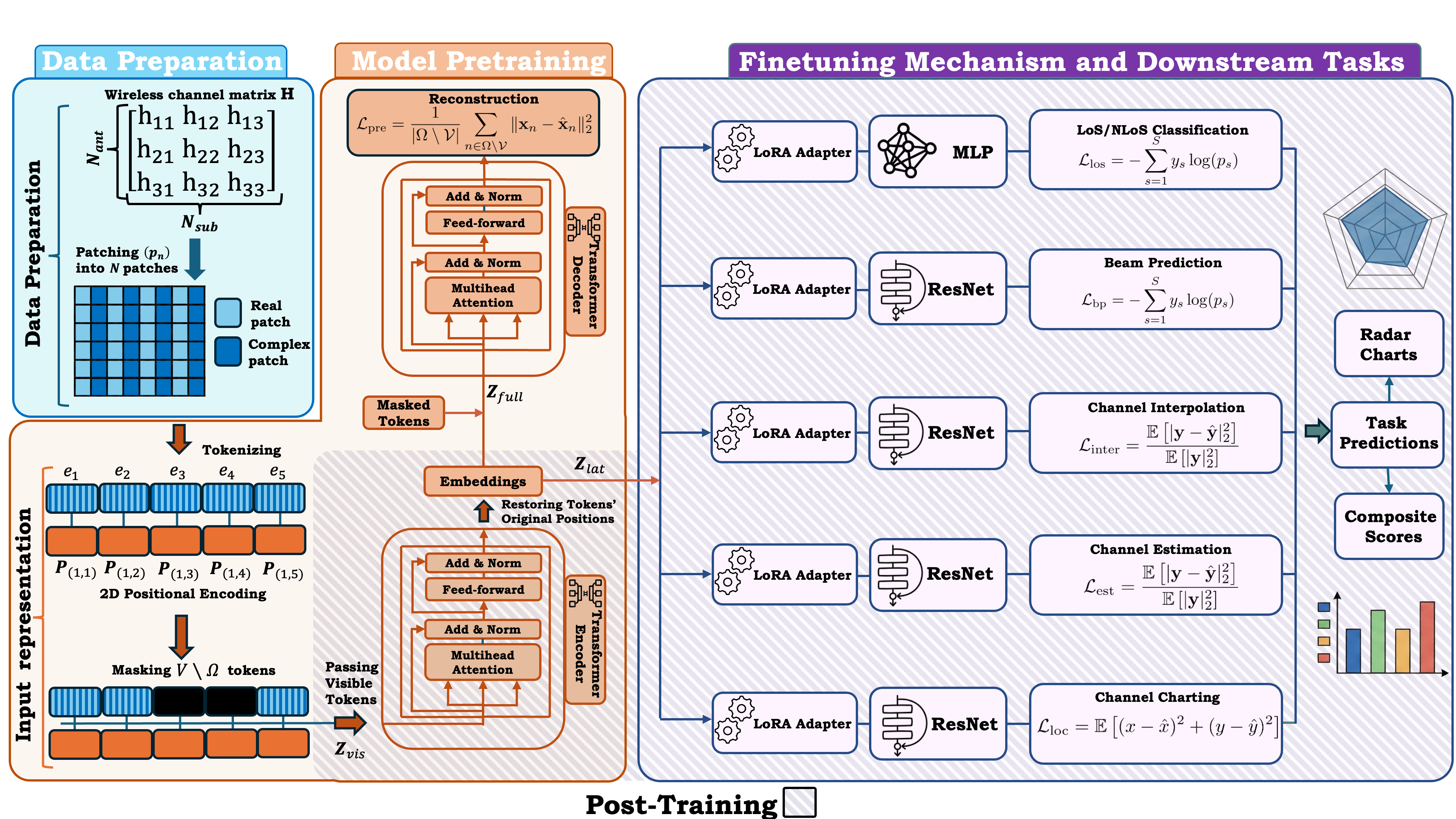}
\caption{Proposed Design Architecture.}
\label{fig:architecture}
\end{figure*}

In this section, we provide the architectural details of each stage in Fig.~\ref{fig:architecture}, focusing on the representation design, transformer backbone, and parameter-efficient task adaptation.

\subsection{Data Preparation: Channel  Tokenization}
\label{subsec:data}

This stage transforms raw complex-valued channel realizations into structured transformer-compatible tokens while preserving the underlying antenna–subcarrier geometry.

\subsubsection{{Grid-Based Patching}}
Unlike standard tokenization methods that flatten the entire channel matrix prior to segmentation, in this work, a grid-based partitioning strategy is adopted as illustrated in the data preparation stage of Fig.~\ref{fig:architecture}. We partition $\mathbf{H}$
into a grid of non-overlapping rectangular patches $\mathbf{p}_n \in \mathbb{C}^{P_h \times P_w}$, defined as
\begin{equation}
\mathbf{p}_n = \mathbf{H}(i_n : i_n + P_h - 1,\; j_n : j_n + P_w - 1),
\end{equation}
where $(i_n, j_n)$ denotes the starting antenna and subcarrier indices of the $n$-th patch. Here, $P_h$ and $P_w$ represent the patch height (spanning antennas) and width (spanning subcarriers), respectively. 
The total number of patches $N$ is given by
\begin{equation}
N = 
\left\lceil \frac{N_{\rm ant}}{P_h} \right\rceil
\left\lceil \frac{N_{\rm sub}}{P_w} \right\rceil.
\end{equation}
This formulation naturally supports \textbf{\textit{variable-sized}} CSI matrices in which changes in $N_{\rm ant}$ or $N_{\rm sub}$ modify only the token count $N$, while the embedding dimension remains fixed. Furthermore, the formulation allows for manual configuration of the patch geometry. For instance, a $4 \times 4$ configuration ($P_h=4, P_w=4$) emphasizes local correlations in both domains, while a $1 \times 16$ configuration ($P_h=1, P_w=16$) captures broader frequency dependencies for a single antenna element.

\subsubsection{Complex Token Representation} 
To process complex-valued channel data efficiently, the real and imaginary components of each patch are stacked together rather than being treated as separate tokens. For a given grid location, the real and imaginary parts of the $\rho = P_h \cdot P_w$ patch elements are flattened and combined into a single raw feature vector $\mathbf{x}_n $:
\begin{equation}
    \mathbf{x}_n = \left[ \text{Re}(\mathbf{p}_n)^\top, \text{Im}(\mathbf{p}_n)^\top \right]^\top \in \mathbb{R}^{L},
    \label{eq:chan_patch}
\end{equation}
where the raw feature dimension is $L = 2 \rho$ and $n \in \{1, \dots, N\}$ represents the sequence index of the patch. By jointly embedding the real and imaginary components into a single token, we maintain a sequence length of $N$ rather than $2N$, significantly reducing the self-attention overhead, which scales quadratically with sequence length ($\mathcal{O}(N^2)$) \cite{ashish_attention}.

\subsection{Input Representation and Masking Mechanism}
This subsection describes the input representation and pretraining framework, including embedding, positional encoding, and masking, used to construct position-aware tokens for self-supervised learning.
\subsubsection{Linear Embedding}
The combined feature vectors $\mathbf{x}_n, \forall n$, are mapped to the latent model dimension $D$ via a trainable linear projection:
\begin{equation}
\mathbf{e}_{n} = \mathbf{W}_{\rm{emb}} \mathbf{x}_n + \mathbf{b}_{\rm{emb}},
\label{proj}
\end{equation}
where $\mathbf{W}_{\rm{emb}} \in \mathbb{R}^{D \times L}$ is the trainable projection matrix, $\mathbf{b}_{\rm{emb}} \in \mathbb{R}^{D}$ is the bias vector, and $\mathbf{e}_{n} \in \mathbb{R}^{D}$ represents the resulting latent token embedding. By retaining the grid structure and jointly embedding the complex components, we ensure that subsequent positional encodings can accurately map each token to its distinct physical location on the antenna-frequency plane.

\subsubsection{2D Positional Encoding}
Unlike natural language sequences where token order is strictly linear, wireless channel data exhibits a fundamental 2D grid structure defined by the spatial (antenna) and spectral (subcarrier) domains. Flattening this grid into a 1D sequence for transformer processing inevitably obscures these orthogonal dependencies. To explicitly preserve the geometric relationships within the MISO-OFDM channel, we employ a 2D learnable PE scheme \cite{dosovitskiy_vit}. 

Although the tokens are processed as a sequence, the underlying 2D grid structure is preserved through patch indexing. Each token $n$ is associated with a spatial–spectral coordinate $(u,v)$ corresponding to its position in the antenna–subcarrier grid.  Because each patch encapsulates a local $P_h \times P_w$ region of the raw channel, the indices track macro-level physical features: $u \in \{1, \dots, H_{\rm grid}\}$ represents the patch row index along the spatial axis (corresponding to a sub-array of adjacent antennas), and $v \in \{1, \dots, W_{\rm grid}\}$ represents the patch column index along the spectral axis (corresponding to a sub-band of adjacent frequency subcarriers). Here, $H_{\rm grid}$ and $W_{\rm grid}$ denote the dynamic spatial and spectral patch dimensions of the current tokenized input.

To encode positional information, we define two learnable embedding matrices:
$\mathbf{E}_{\rm ant} \in \mathbb{R}^{H_{\rm grid} \times d_{\rm ant}}, \quad
\mathbf{E}_{\rm sub} \in \mathbb{R}^{W_{\rm grid} \times d_{\rm sub}},$ 
where $d_{\rm ant} + d_{\rm sub} = D$. These matrices act as lookup tables that map discrete spatial and spectral indices to continuous latent representations. Unlike fixed sinusoidal encodings, $\mathbf{E}_{\rm ant}$ and $\mathbf{E}_{\rm sub}$ are fully learnable parameters optimized during training, allowing the model to adapt positional representations to the statistical structure of wireless channels.

Formally, for a given coordinate $(u,v)$, the positional components are obtained as:
\begin{equation}
\mathbf{e}_{\rm ant}^{(u)} = \mathbf{E}_{\rm ant}[u,:] \in \mathbb{R}^{d_{\rm ant}}, \quad
\mathbf{e}_{\rm sub}^{(v)} = \mathbf{E}_{\rm sub}[v,:] \in \mathbb{R}^{d_{\rm sub}}.
\end{equation}
This can equivalently be interpreted as a linear embedding of one-hot encoded indices.

The full positional encoding is constructed via a separable concatenation:
\begin{equation}
\mathbf{P}_{(u,v)} =
\begin{bmatrix}
\mathbf{e}_{\rm ant}^{(u)} \\
\mathbf{e}_{\rm sub}^{(v)}
\end{bmatrix}
\in \mathbb{R}^{D}.
\label{2d_pe}
\end{equation}
This separable formulation introduces an explicit inductive bias that decouples spatial and spectral representations, enabling the model to independently capture antenna-domain correlations (e.g., beamforming patterns) and frequency-domain variations (e.g., channel selectivity). 
The positional encoding is then added to the token embedding:
\begin{equation}
\mathbf{z}_n = \mathbf{e}_n + \mathbf{P}_{(u,v)},
\label{2d_pe_add}
\end{equation}
yielding position-aware representations that jointly encode channel content and spatial–spectral location within a unified latent space.

\subsubsection{{Masking Strategy}} 
Following tokenization and 2D positional embedding, we apply the random masking  to facilitate the self-supervised reconstruction objective. Let $\Omega = \{1, \dots, N\}$ denote the set of all patch sequence indices. The masking operator uniformly samples a subset of visible indices $\mathcal{V} \subset \Omega$ such that $|\mathcal{V}| = \lfloor (1 - m) N \rfloor$ and $m$ is the masking ratio. The indices in the complement set $\Omega \setminus \mathcal{V}$ are entirely discarded from the encoder's view. 

This selective processing represents a critical efficiency feature of the architecture. Unlike standard masked language models that process full sequences populated with placeholder tokens, our encoder operates exclusively on the reduced visible subset. The final input to the transformer encoder, $\mathbf{Z}_{\rm{vis}} \in \mathbb{R}^{|\mathcal{V}| \times D}$, is constructed by gathering only the position-aware embeddings $\mathbf{z}_n$ corresponding to the visible indices:
\begin{equation}
\mathbf{Z}_{\rm{vis}} = \{ \mathbf{z}_n \mid n \in \mathcal{V} \}.
\end{equation}
Because the physical geometry of the channel is already explicitly encoded within each token $\mathbf{z}_n$ via the 2D positional encoding, the sequence length can be drastically reduced without losing the spatial-spectral topology of the data.

\subsection{Transformer Backbone}
\label{subsec:backbone}

The backbone follows MAE design composed of (i) a transformer encoder that produces latent channel representations from visible tokens, and (ii) a lightweight decoder used only during pretraining to reconstruct masked tokens. After pretraining, the decoder is removed and the encoder is retained for downstream adaptation.

\subsubsection{Encoder Architecture} 
The encoder consists of $L_{\rm enc}$ stacked transformer layers, operating specifically on the visible subset of the input sequence. Let $\mathbf{Z}_{\ell-1}$ denote the input to layer $\ell \in \{1, \dots, L_{\rm enc}\}$, where the input to the first layer is the visible patch sequence ($\mathbf{Z}_{\rm vis}$), and the input to any subsequent layer is the output of the preceding one. Each layer processes this representation through two primary sub-layers: multi-head self-attention (MHSA) and a feed-forward network (FFN). Residual connections and layer normalization (LN) are employed to facilitate stable deep learning.

\paragraph{\textbf{Multi-Head Self-Attention (MHSA)}}

To jointly model heterogeneous spatial–frequency dependencies, we employ MHSA with $H$ parallel attention heads. For each head $h \in \{1,\dots,H\}$, each sub-layer follows the pre-normalized convention, in which an LN is applied to the input before projection. Letting
$\widetilde{\mathbf{Z}}_{\ell-1}=\mathrm{LN}(\mathbf{Z}_{\ell-1})$ denote the normalized layer input, the per-head query, key, and value
projections are computed as
\begin{equation}
\mathbf{Q}_h = \widetilde{\mathbf{Z}}_{\ell-1}\mathbf{W}_Q^{h}, \quad
\mathbf{K}_h = \widetilde{\mathbf{Z}}_{\ell-1}\mathbf{W}_K^{h}, \quad
\mathbf{V}_h = \widetilde{\mathbf{Z}}_{\ell-1}\mathbf{W}_V^{h},
\end{equation}
where $\mathbf{W}_Q^{h}, \mathbf{W}_K^{h}, \mathbf{W}_V^{h} \in \mathbb{R}^{D \times d_h}$ are learnable projection matrices and $d_h = D/H$ denotes the per-head feature dimension. The attention weights for head $h$ are computed using scaled dot-product attention
\begin{equation}
\mathbf{A}_h = \mathrm{softmax}\!\left(\frac{\mathbf{Q}_h\mathbf{K}_h^\top}{\sqrt{d_h}}\right),
\end{equation}
which captures pairwise token interactions exclusively within the visible subset. The corresponding head output is
\begin{equation}
\mathrm{head}_h = \mathbf{A}_h \mathbf{V}_h 
\in \mathbb{R}^{|\mathcal{V}| \times d_h}.
\end{equation}
The outputs of all $H$ heads are concatenated and linearly projected to restore the latent model dimension
\begin{equation}
\mathrm{MHSA}(\mathbf{Z}_{\ell-1}) =
\mathrm{Concat}(\mathrm{head}_1,\dots,\mathrm{head}_H)\mathbf{W}_O,
\end{equation}
where $\mathbf{W}_O \in \mathbb{R}^{D \times D}$ is the output projection matrix. This formulation enables the encoder to attend to channel correlations across multiple representation subspaces simultaneously, allowing distinct heads to specialize in complementary spatial and frequency interaction patterns while preserving global contextual aggregation across the visible tokens.

\paragraph{\textbf{Position-wise FFN and LN}}

Given an intermediate token representation $\mathbf{z} \in \mathbb{R}^D$, LN is applied along the feature dimension independently for each token:
\begin{equation}
\mathrm{LN}(\mathbf{z}) 
= \frac{\mathbf{z}-\mu}{\sqrt{\sigma^2+\epsilon}} 
\odot \boldsymbol{\gamma} + \boldsymbol{\beta},
\end{equation}
where $\mu$ and $\sigma^2$ are the mean and variance computed over the $D$ features of the token, $\epsilon$ ensures numerical stability, and $\boldsymbol{\gamma},\boldsymbol{\beta} \in \mathbb{R}^D$ are learnable affine parameters. The position-wise FFN is then applied identically and independently to each token:
\begin{equation}
\mathrm{FFN}(\mathbf{z})
= \phi\!\left(\mathrm{LN}(\mathbf{z})\,\mathbf{W}_1 + \mathbf{b}_1\right)\mathbf{W}_2 + \mathbf{b}_2 ,
\end{equation}
where $\mathbf{W}_1 \in \mathbb{R}^{D \times D_{\rm ff}}$ and $\mathbf{W}_2 \in \mathbb{R}^{D_{\rm ff} \times D}$ are the weight matrices, $\mathbf{b}_1 \in \mathbb{R}^{D_{\rm ff}}$ and $\mathbf{b}_2 \in \mathbb{R}^{D}$ are the biases, and $\phi(\cdot)$ denotes a general non-linear activation function; and $D_{\rm ff}$ defines the intermediate hidden expansion dimension. To ensure stable optimization in deep architectures, the complete transformer layer $\ell$ follows the pre-normalized residual structure
\begin{align}
\mathbf{Z}'_{\ell} &= \mathbf{Z}_{\ell-1} + \mathrm{MHSA}(\mathbf{Z}_{\ell-1}), \\
\mathbf{Z}_{\ell}  &= \mathbf{Z}'_{\ell}   + \mathrm{FFN}(\mathbf{Z}'_{\ell}),
\end{align}
ensuring stable optimization while preserving token-wise feature transformations.

\subsubsection{Decoder and Reconstruction}
The decoder is a lightweight transformer consisting of $L_{\rm dec}$ layers (typically $L_{\rm dec} < L_{\rm enc}$) designed to reconstruct the full channel matrix from the latent representation. Unlike the encoder, which processes only the visible subset, the decoder operates on the full sequence $\Omega$ of length $N$. The reconstruction process involves sequence restoration, positional encoding injection, and feature projection, as detailed below.

\paragraph{\textbf{Sequence Restoration and Mask Tokens}}

First, we recover the complete patch sequence by restoring the original ordering of the visible patches and inserting learnable mask tokens for the previously discarded positions. We initialize a full sequence representation $\mathbf{Z}_{\rm full} \in \mathbb{R}^{N \times D}$. For each token index $n$ in the grid
\begin{equation}
    \mathbf{z}_{n}^{\rm full} = 
    \begin{cases} 
        \mathbf{z}_{n}^{\rm enc} & n \in \mathcal{V}, \\
        \mathbf{e}_{\rm mask} & n \notin \mathcal{V},
    \end{cases}
\end{equation}
where $\mathbf{z}_{n}^{\rm enc}$ is the corresponding encoded latent vector output by the encoder, and $\mathbf{e}_{\rm mask} \in \mathbb{R}^D$ is a shared, learnable mask token.

\paragraph{\textbf{Positional Encoding Restoration}}
Since the mask tokens $\mathbf{e}_{\rm mask}$ are identical for all missing positions, they possess no inherent spatial or spectral information. To restore geometric context across the 2D plane, we add a full set of decoder positional encodings $\mathbf{P}_{\rm dec} \in \mathbb{R}^{N \times D}$ to the restored sequence, yielding the final position-aware decoder input sequence, denoted as $\mathbf{Z}_{\rm in} \in \mathbb{R}^{N \times D}$:
\begin{equation}
    \mathbf{Z}_{\rm in} = \mathbf{Z}_{\rm full} + \mathbf{P}_{\rm dec}.
\end{equation}
This addition is critical; it ensures the decoder can uniquely identify the location of every masked patch and accurately propagate physical channel information from the surrounding visible anchors.

\paragraph{\textbf{Decoder Layers}}
The complete input sequence $\mathbf{Z}_{\rm in}$ is then processed through $L_{\rm dec}$ transformer layers. These layers mirror the encoder architecture (employing MHSA and FFN), but are functionally shallower. The global self-attention mechanism enables the mask tokens to attend to the visible patches, effectively interpolating the missing channel state information based on the learned spatial-frequency context.

\paragraph{\textbf{Final Projection and Reconstruction}}
After passing through the decoder layers, the output feature matrix $\mathbf{Z}_{\rm out} \in \mathbb{R}^{N \times D}$ is projected back to the original raw element space via a linear projection layer:
\begin{equation}
    \hat{\mathbf{X}} = \mathbf{Z}_{\rm out}\mathbf{W}_{\rm out} + \mathbf{b}_{\rm out},
\end{equation}
where $\mathbf{W}_{\rm out} \in \mathbb{R}^{D \times L}$ and $\mathbf{b}_{\rm out} \in \mathbb{R}^{L}$ are the output projection parameters, and $L$ is the flattened complex patch dimension established during the initial tokenization. Finally, to recover the predicted complex-valued channel matrix $\hat{\mathbf{H}}$, the output vectors in $\hat{\mathbf{X}}$ are split into their real and imaginary components, unflattened, and recombined according to the original $\mathbb{C}^{N_{\rm ant} \times N_{\rm sub}}$ spatial-spectral grid.

\section{ Model Training and Fine Tuning \\ for Downstream Wireless Tasks}
\subsection{Model Pretraining}
The model is pre-trained via self-supervised learning to minimize the reconstruction error. We employ the mean squared error (MSE) as the objective function. Crucially, the loss is calculated only on the masked patches, forcing the model to infer the missing channel structures from global context rather than memorizing the visible input. Given the ground truth patch vectors $\mathbf{x}_k$ and the reconstructed output vectors $\hat{\mathbf{x}}_k$ from the decoder, the loss function can be expressed as
\begin{equation}
\mathcal{L}_{\rm{pre}} = \frac{1}{|\Omega \setminus \mathcal{V}|} \sum_{n \in \Omega \setminus \mathcal{V}} \| \mathbf{x}_n - \hat{\mathbf{x}}_n \|_2^2,
\label{eq:pretrain}
\end{equation}
where $\Omega \setminus \mathcal{V}$ represents the set of masked patch indices (the complement to the visible subset), $|\Omega \setminus \mathcal{V}|$ is the total number of masked patches, and $\mathbf{x}_n, \hat{\mathbf{x}}_n \in \mathbb{R}^L$ are the original and reconstructed flattened feature vectors for the $n$-th patch, respectively. By strictly optimizing over the unobserved regions, this objective drives the encoder to learn robust, scenario-agnostic latent representations of the physical wireless channel.


A defining feature of the MAE encoder is its \textit{efficiency}. Because the self-attention mechanism computes pairwise interactions exclusively among the visible tokens (discarding the masked tokens entirely), the computational complexity is significantly reduced. Standard attention scales quadratically with the full sequence length, $\mathcal{O}(N^2)$. However, by operating solely on the unmasked subset $\mathcal{V}$, the complexity becomes $\mathcal{O}(|\mathcal{V}|^2) = \mathcal{O}((1-m)^2 N^2)$. For a high masking fraction (e.g., $m = 0.75$), the encoder processes only 25\% of the total patches. This results in a drastic, quadratic reduction in memory and compute usage, allowing the model to scale efficiently to high-resolution channel grids and larger batch sizes without the prohibitive costs associated with standard transformers.



\subsection{Finetuning Mechanism and Downstream Tasks}
To adapt the pre-trained foundation model to specific downstream tasks, we employ a parameter-efficient fine-tuning (PEFT) strategy that transitions the architecture from a reconstructive objective to a discriminative one, as outlined in Fig.~\ref{fig:architecture}.

\subsubsection{Backbone Adaptation} 
The transition to the fine-tuning stage involves structural modifications to the foundation model. We discard the lightweight pre-training decoder and the random masking operator, retaining only the pre-trained WALoMA encoder $f_\theta$ to serve as the generic feature backbone. 

Unlike the pretraining phase, the encoder now processes the full, unmasked channel sequence. Let $\mathbf{Z}_{\rm seq} \in \mathbb{R}^{N \times D}$ denote the complete input matrix formed by gathering all position-aware patch embeddings $\mathbf{z}_n$ for $n \in \Omega$. The encoder $f_\theta$ processes this full sequence to produce a dense, globally contextualized latent representation $\mathbf{Z}_{\rm lat} = f_\theta(\mathbf{Z}_{\rm seq})$, where the final output matrix retains the dimensions $\mathbf{Z}_{\rm lat} \in \mathbb{R}^{N \times D}$.

\subsubsection{Low-Rank Adaptation} 
To achieve parameter-efficient adaptation and avoid full fine-tuning of the transformer backbone, we adopt LoRA, which reduces the number of trainable parameters while mitigating catastrophic forgetting. Specifically, we re-parameterize the encoder's weights $\theta$. Rather than fine-tuning all parameters of the WALoMA encoder $f_\theta(\cdot)$, we freeze the pre-trained weights $\mathbf{W}_0$ and inject trainable rank decomposition matrices into the linear layers of the transformer blocks. For a pre-trained weight matrix $\mathbf{W}_0 \in \mathbb{R}^{d \times n}$, the weight update is parameterized as
\begin{equation}
    \mathbf{W} = \mathbf{W}_0 + \frac{\alpha}{r} \mathbf{B}\mathbf{A},
\end{equation}
where $\mathbf{B} \in \mathbb{R}^{d \times r}$ and $\mathbf{A} \in \mathbb{R}^{r \times n}$ are low-rank matrices with $r \ll \min(d, n)$. 

\subsubsection{Downstream Tasks and Adaptation Heads}
Following the formulation in Section III, for each task $k \in \{1, \dots, K\}$, a specialized evaluation head $g_\phi^{(k)}(\cdot)$ is attached to the backbone. These heads are designed to map the high-dimensional features in $\mathcal{Z}$ to specific wireless metrics. For each task, a specialized evaluation head is attached to the backbone, corresponding to the multi-task outputs of Fig.~\ref{fig:architecture}.

\textbf{ Task 1: LoS/NLoS Classification:} A classical binary classification task in wireless communication to determine if the UE has a LoS path to the BS. Given that LoS/NLoS detection is a comparatively low-complexity binary classification task, we utilize a lightweight multilayer perceptron (MLP) head $g_\phi^{(1)}(\cdot)$, where the patch embeddings $\mathbf{Z}_{\rm lat}$ are flattened and passed through a hidden layer with batch normalization and ReLU activation. A final linear layer maps these features to 2 class logits. The loss function is a cross-entropy loss defined as 
\begin{equation}
\mathcal{L}_{\rm{los}} = - \sum_{s=1}^{S} y_{s} \log(p_{s}),
\label{eq:los_loss}
\end{equation}
where $S$ is the total number of classes, $y_s \in \{0, 1\}$ is the ground-truth binary indicator for class $s$, and $p_s$ is the predicted probability for class $s$ derived from the softmax-normalized logits.

\textbf{Task 2: Beam Prediction:} This task identifies the optimal beam index among $N_B$ beams. We employ a ResNet-inspired 2D CNN head $g_\phi^{(2)}(\cdot)$, chosen for its ability to capture complex spatial patterns in the channel grid without the vanishing gradient issues typical of deep networks. The input embeddings are first projected via a stem convolution to $32$ channels. This is followed by three residual stages with increasing widths ($32 \rightarrow 64 \rightarrow 128$). A global average pooling layer collapses the spatial dimensions before a final linear classifier with a 0.3 dropout rate maps the features to the beam logits. The loss function $\mathcal{L}_{\rm{bp}}$ is a cross-entropy loss defined in  \eqref{eq:los_loss} for $S=N_B$.

\textbf{Task 3: Channel Interpolation:} Reconstructing the full dense channel response from the latent features. The tokens are reshaped into a 2D grid and passed through a projection stem (128 channels). The model then distills information through progressive residual blocks ($128 \rightarrow 64 \rightarrow 32$). The final output is reshaped and processed by a patch reconstructor to recover the original dimensions. The objective is to minimize the normalized mean squared error (NMSE):
\begin{equation}
\mathcal{L}_\text{inter} = \frac{\mathbb{E} \left[ | \mathbf{y} - \hat{\mathbf{y}} |_2^2 \right]}{\mathbb{E} \left[ | \mathbf{y} |_2^2 \right]},
\label{eq:nmse_linear}
\end{equation}
where $\mathbf{y}$ represents the ground-truth channel vector and $\hat{\mathbf{y}} = g_{\phi}^{(3)}(\mathbf{Z}_{\rm lat})$ is the predicted vector. To facilitate a comparative analysis across the varying noise regimes and to align with the challenge's evaluation criteria, the performance is reported in dB as
\begin{equation}
\text{NMSE}_{\text{dB}} = 10 \log{10} \left( \mathcal{L}_\text{inter} \right).
\label{eq:nmse_db}
\end{equation}
The final composite performance score is calculated by mapping these $\text{NMSE}_{\text{dB}}$ values against task-specific limits $[\text{NMSE}_{\text{min}}, \text{NMSE}_{\text{max}}]$, ensuring that the model's accuracy is benchmarked relative to the physical limits of the wireless environment.

\textbf{Task 4: Channel Estimation:} Denoising and estimating accurate CSI from noisy observations. Task head $g_\phi^{(4)}(\cdot)$ utilizes a 2-layer convolutional stack ($128 \rightarrow 64$ channels) with an explicit residual shortcut from the input grid to the final processed features to facilitate gradient flow. The output is projected to the patch element space and reconstructed into a full CSI matrix using the NMSE loss $\mathcal{L}_\text{est}$ and scores defined in  \eqref{eq:nmse_linear} and  \eqref{eq:nmse_db}.

\textbf{Task 5: Channel Charting:} We perform unsupervised manifold learning to regress 2D spatial coordinates. The head $g_\phi^{(5)}(\cdot)$ adopts a funnel architecture similar to Task 3. After global pooling, a regressor MLP (32 $\rightarrow$ 16 $\rightarrow$ 2) predicts the continuous coordinates $(x, y)$. The objective is the MSE in the spatial domain:
\begin{equation}
\mathcal{L}_{\text{loc}} = \mathbb{E} \left[ (x - \hat{x})^2 + (y - \hat{y})^2 \right].
\end{equation}

\subsubsection{Downstream Inference and Evaluation}
In the final stage, the adapted encoder $f_\theta(\cdot)$ processes the input channel data to generate context-aware latent embeddings $\mathbf{Z}_{\rm lat} \in \mathbb{R}^{N \times D}$, which are subsequently passed to the task-specific heads $\{g_\phi^{(1)}, \dots, g_\phi^{(T)}\}$. To rigorously assess the model's generalization capabilities, we utilize task-appropriate metrics: the weighted F1-Score for classification tasks (Tasks 1 and 2), NMSE for CSI reconstruction (Tasks 3 and 4), and MSE in meters for channel charting (Task 5). Performance is visualized via a radar chart comparing the optimized model against baseline model (LWM).

\begin{table}[t!]
\caption{Simulation and model parameters}
\label{tab:params}
\centering
\footnotesize
\setlength{\tabcolsep}{3pt} 
\begin{tabular}{|l|c|}
\hline
\textbf{Parameter} & \textbf{Value} \\
\hline
\multicolumn{2}{|c|}{\textit{DeepMIMO Environment}} \\
\hline
Scenarios & \begin{tabular}{@{}c@{}}City (0-19), O1, \\ ASU Campus, Boston5G\end{tabular} \\
Frequency & 3.5 GHz \\
BS Antennas ($N_{ant}$) & $\{8, 16, 32, 64, 128\}$ \\
Subcarriers ($N_{sub}$) & $\{32, 64, \ldots, 1024\}$ \\
Number of beams ($N_B$) & $\{8,64\}$ \\
\hline
\multicolumn{2}{|c|}{\textit{Model Architecture}} \\
\hline
Embed. Dim ($D$) & 64 \\
Encoder layers & 12 \\
Decoder layers & 4 \\
Encoder attention heads & 12 \\
Decoder attention heads & 4 \\
Patch Size & $4 \times 4$ \\
Masking ratio $m$ & $0.75$ \\
Dropout & 0.1 \\
Encoder parameters & $\approx$ 0.6M \\
Decoder parameters & $\approx$ 0.2M \\
\hline
\multicolumn{2}{|c|}{\textit{Pretraining}} \\
\hline
Pretraining samples & $\approx$ 2.45M \\
Epochs & 1000 \\
Batch Size & 768 \\
Optimizer & AdamW \\
Learning Rate & $3 \times 10^{-4}$ \\
Weight Decay & 0.05 \\
\hline
\end{tabular}
\end{table}

\section{Numerical Results and Discussion}

\begin{figure}[ht!]
\centering
\includegraphics[width=\columnwidth]{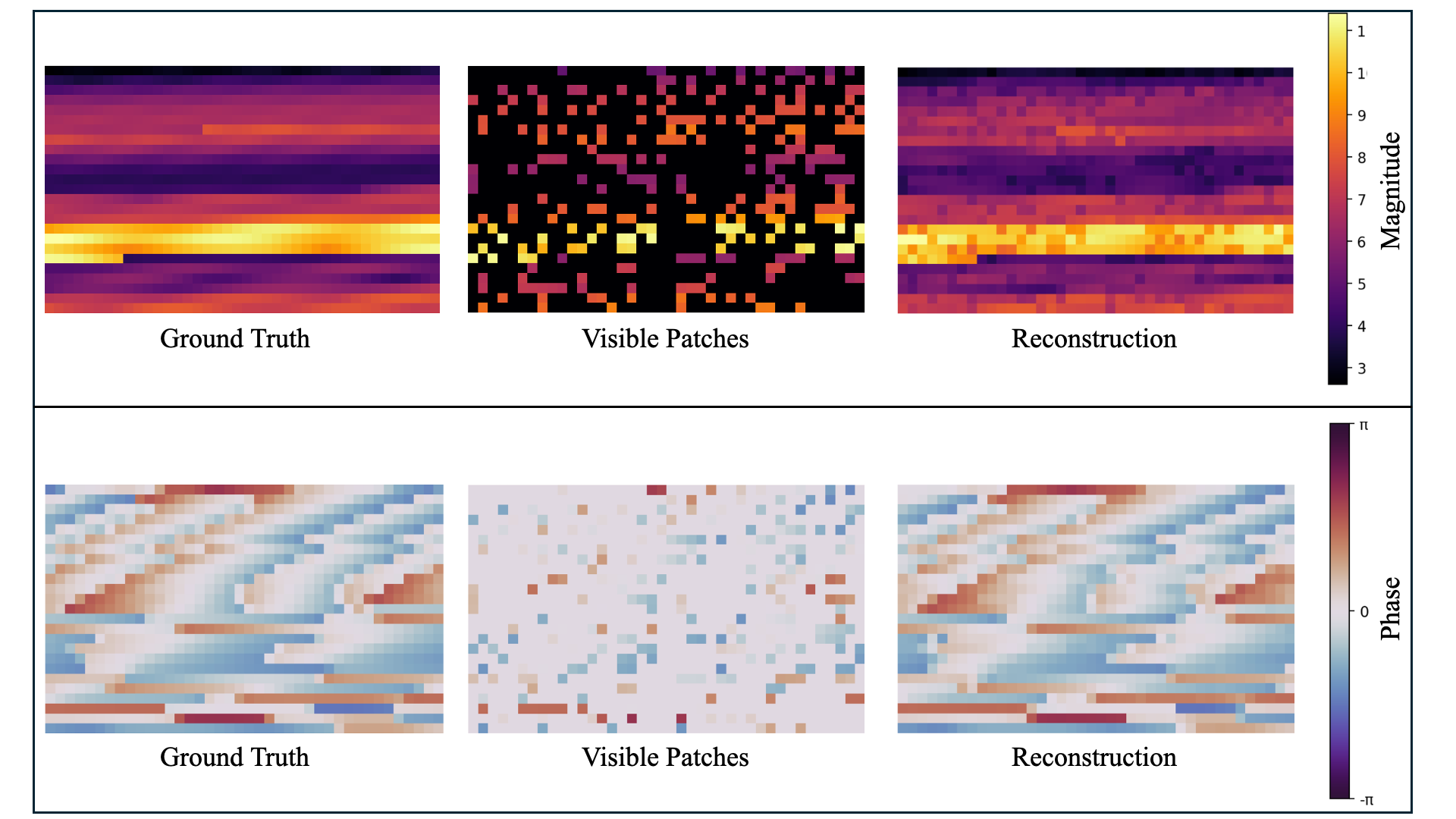}
\caption{Training validation - comparison between ground-truth channels to reconstructed ones.}
\label{fig:pretraining}
\end{figure} 

\begin{figure}[ht!]
\centering
\includegraphics[width=0.75\columnwidth]{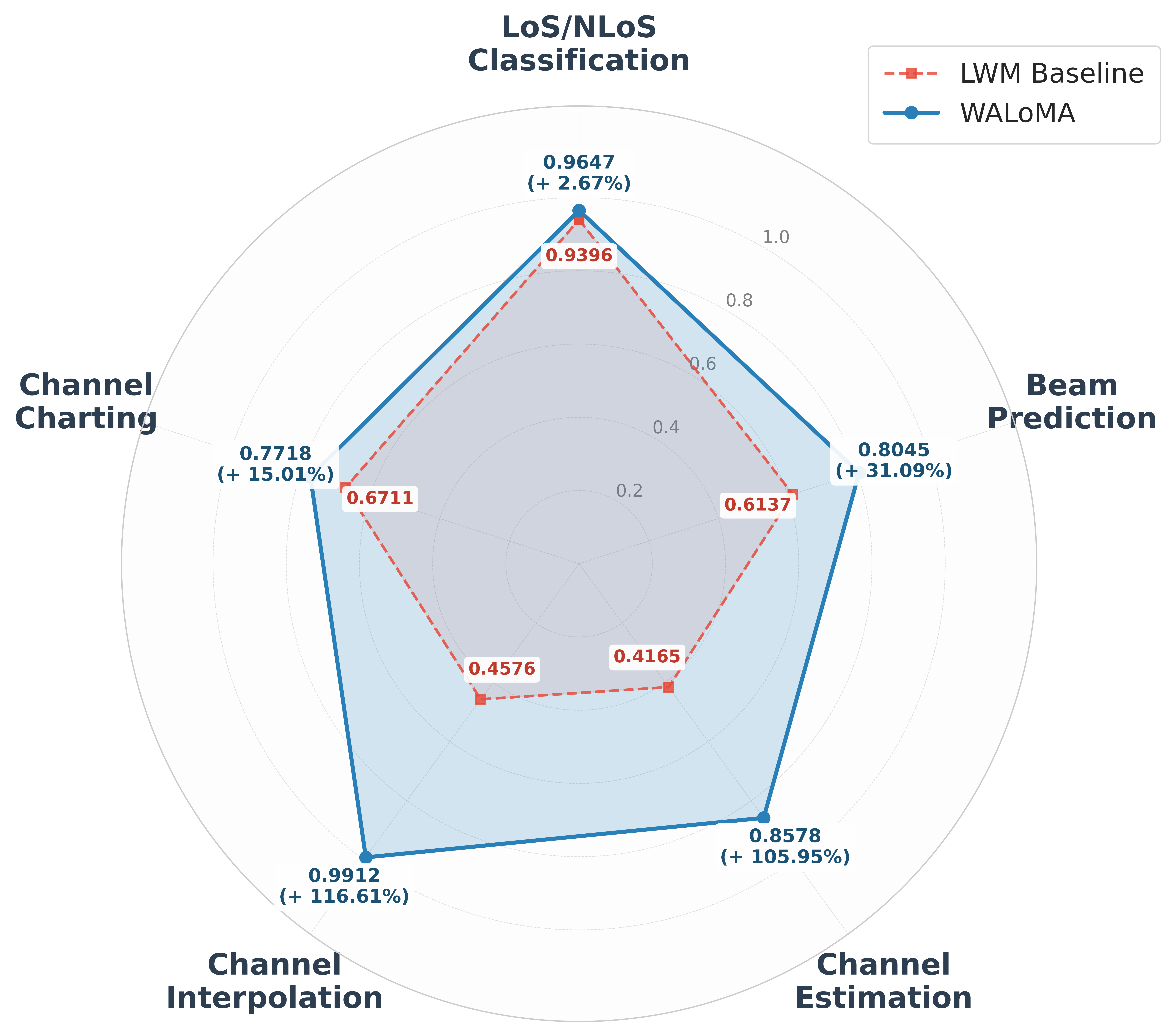}
\caption{Comparison of Composite Performance Scores: WALoMA vs. LWM Baseline}
\label{fig:chart}
\end{figure} 

\begin{table}[htbp]
\small
\centering
\caption{Dataset Partitioning for Downstream Evaluation}
\label{tab:dataset_partitioning}
\begin{tabular}{l c c c}
\hline
\textbf{Task}  & \textbf{Training} & \textbf{Validation} & \textbf{Test} \\
\hline
LoS/NLoS  & 6 & 500 & 1500 \\
Beam Prediction         & 1000 & 500 & 1500 \\
Channel Interpolation   & 300 & 500 & 1500 \\
Channel Estimation      & 250 & 500 & 1500 \\
Channel Charting        & 450 & 500 & 1500 \\
\hline
\end{tabular}
\end{table}

In this section, we evaluate the efficacy of the proposed foundation model and validate its generalization capabilities across diverse downstream wireless tasks. The default simulation and architectural parameters are listed in Table \ref{tab:params}, unless specified otherwise. All model pre-training was conducted using six NVIDIA Tesla V100 GPUs, while the downstream task adaptation and evaluation were performed on a single V100 GPU.
\subsubsection{Deployment Scenarios and Variability}

To encourage scenario-agnostic representation learning, we utilize large-scale channel realizations generated by the DeepMIMO \cite{alkhateeb_deepmimo} ray-tracing framework. The dataset spans heterogeneous propagation environments with diverse geometric and blockage characteristics.

Specifically, training and evaluation are conducted across:

\begin{itemize}
    \item \textbf{Urban city scenarios (City 0--19):} 20 distinct metropolitan layouts with substantial variability in building density and height.
    \item \textbf{Campus and specialized scenarios:} ASU Campus (mixed vegetation and structures), O1/O1b (controlled outdoor blockage), and Boston5G (dense urban deployment).
\end{itemize}

This diversity exposes the model to varying spatial correlation structures, delay spreads, and angular distributions, promoting generalization across deployment conditions. For the downstream evaluation, the dataset is partitioned into specific training, validation, and test sets to ensure rigorous assessment, as summarized in Table~\ref{tab:dataset_partitioning}.

\subsection{Reconstruction Performance}
Fig. \ref{fig:pretraining} demonstrates the model's capability for high-fidelity reconstruction of both the magnitude (top row) and phase (bottom row) components of the wireless channel. The visual progression illustrates the ground truth channel matrices (left), the highly sparse visible patches provided to the network (middle), and the complete reconstructed outputs (right). 

Notably, the visible patches within the reconstructed matrices differ slightly from their ground-truth counterparts. Rather than a flaw, this variance is a fundamental characteristic of the MAE architecture. The final output is not a mere identity mapping of the visible input; instead, the entire spatial-spectral sequence undergoes a rigorous compression and decompression process. During self-supervised pre-training, the encoder heavily compresses the high-dimensional channel data into a lower-dimensional latent space, and the decoder subsequently generates the full channel response entirely from scratch.

Consequently, this inherent reconstruction loss is directly governed by the dimensionality of the embedding vector. A constrained latent space acts as an information bottleneck, forcing the model to distill the most salient, generalized physical features of the propagation environment rather than simply memorizing exact input values. This compression behavior is crucial to the framework's overall efficacy, as these highly robust latent embeddings are precisely what feed into the adaptation and inference heads, ensuring efficient downstream multi-task performance without overfitting to the pretraining data.

\subsection{Comparative Evaluation Across Downstream Tasks}

After self-supervised pretraining, the decoder is discarded, and the encoder is retained as a pretrained feature extractor for efficient adaptation to downstream tasks. As illustrated in Fig.~\ref{fig:chart}, the proposed model demonstrates a substantial advantage over the LWM baseline across five critical evaluations. Most notably, the architecture achieves a score of 0.8045 in beam prediction, a 31.09\% improvement over the LWM baseline (0.6137). Even more dramatic leaps are observed in dense spatial reconstruction, with both channel estimation (from 0.4165 to 0.8578) and channel interpolation (from 0.4576 to 0.9912) experiencing over a 100\% increase in accuracy.

Conversely, the LoS/NLoS classification task exhibits a much narrower margin of improvement, shifting only 2.67\% (from 0.9396 to 0.9647). This plateau is structurally expected; determining LoS/NLoS presence is fundamentally a low-complexity binary classification problem that relies primarily on macroscopic power thresholding. Because it does not inherently require deep comprehension of complex spatial and spectral correlations, the rich geometric representations captured by our advanced positional encoding do not yield the massive comparative advantage seen in analytically demanding regression tasks. Ultimately, the marked overall increase in performance across the framework is directly attributed to the transition to an MAE integrated with 2D PE, alongside significant architectural upgrades to the downstream convolutional task heads.

\begin{table}[ht]
\centering
\small
\caption{Ablation study: PE vs. LoRA efficiency.}
\label{tab:ablation_final}
\begin{tabularx}{\columnwidth}{@{} l >{\centering\arraybackslash}X >{\centering\arraybackslash}X | >{\centering\arraybackslash}X >{\centering\arraybackslash}X @{}}
\toprule
\textbf{Task} & \textbf{1D PE} & \textbf{2D PE} & \textbf{LoRA} & \textbf{Param \%} \\ \midrule
LoS/NLoS              & 96.81         & \textbf{96.87}         & 96.47        & 12.83             \\
Beam Prediction              & 79.53         & \textbf{80.51} & 80.45        & 11.27             \\
Channel Interpolation              & 59.48         & \textbf{79.44}         & \textbf{85.78} & 12.84             \\
Channel Estimation              & 97.28         & \textbf{98.68}         & \textbf{99.12} & 16.88             \\
Channel Charting              & 66.08         & \textbf{75.26}         & \textbf{77.18} & 19.59             \\ \midrule
\textbf{Composite} & 79.84         & \textbf{86.15}         & \textbf{87.80} & 14.68               \\ \bottomrule
\end{tabularx}
\label{tab:ablation}
\end{table}
\subsection{Ablation Study}
The ablation study results, summarized in Table \ref{tab:ablation}, quantify the individual and cumulative contributions of the proposed architectural enhancements, demonstrating a significant performance trajectory. The transition from 1D to 2D PE provides the first major performance leap, particularly in spatially-sensitive tasks such as channel interpolation and channel charting where scores increased from 59.48\% to 79.44\% and 66.08\% to 75.26\%, respectively, confirming that explicitly modeling the two-dimensional grid structure of the channel is essential for capturing the physical correlations between antennas and subcarriers. Further gains are realized through the integration of LoRA, which pushes the composite score to 87.80\% while only requiring the training of a small fraction of the total model parameters, averaging just 14.68\% across all tasks. The effectiveness of LoRA is most evident in channel estimation, which reaches a peak score of 99.12\%, validating that LoRA successfully preserves the rich latent embeddings learned by the MAE. Consequently, the proposed configuration achieves superior performance across most metrics with substantially reduced resource consumption, establishing a highly scalable framework for advanced wireless communication tasks.

\begin{table}[t]
\centering
\caption{Task Performance (\%) Across Training Budgets.}
\label{tab:training_budget_comparison}
\footnotesize
\begin{tabularx}{0.45\textwidth}{@{} l l *{5}{>{\centering\arraybackslash}X} @{}}
\toprule
\textbf{Model} & \textbf{Task} & \textbf{10\%} & \textbf{25\%} & \textbf{50\%} & \textbf{75\%} & \textbf{100\%} \\ \midrule

\textbf{LWM} & LoS/NLoS & N/A & N/A & 31.936 & 29.019 & 93.962 \\
             & Beam Prediction & 14.305 & 32.876 & 44.405 & 58.962 & 61.037 \\
             & Channel Interpolation & 14.209 & 15.051 & 28.294 & 33.446 & 41.650 \\
             & Channel Estimation & 14.233 & 31.011 & 34.423 & 44.993 & 45.760 \\
             & Channel Charting & 33.575 & 33.361 & 50.319 & 62.717 & 67.110 \\ \midrule

\textbf{Raw} & LoS/NLoS & N/A & N/A & 22.477 & 23.080 & 69.738 \\
             & Beam Prediction & 16.936 & 32.408 & 47.146 & 58.079 & 63.846 \\
             & Channel Interpolation & 21.615 & 24.544 & 26.127 & 27.029 & 27.477 \\
             & Channel Estimation & \textbf{40.885} & \textbf{70.426} & \textbf{73.272} & 75.493 & 75.868 \\
             & Channel Charting & 15.694 & 25.758 & 28.865 & 33.983 & 38.951 \\ \midrule

\textbf{Proposed} & LoS/NLoS & N/A & N/A & \textbf{37.716} & \textbf{37.716} & \textbf{96.471} \\
              & Beam Prediction & \textbf{19.127} & \textbf{51.500} & \textbf{65.244} & \textbf{73.805} & \textbf{80.450} \\
              & Channel Interpolation & \textbf{24.869} & \textbf{47.778} & \textbf{58.557} & \textbf{80.799} & \textbf{85.784} \\
              & Channel Estimation & 19.050 & 49.162 & 55.750 & \textbf{88.726} & \textbf{99.118} \\
              & Channel Charting & \textbf{33.725} & \textbf{40.722} & \textbf{60.762} & \textbf{72.941} & \textbf{77.182} \\ \bottomrule
\end{tabularx}
\label{tab:train_budget}
\end{table}
\subsection{Task Performance Scores Across Training Budgets.}
The performance of the proposed model across varying training data availability is shown in Table  \ref{tab:train_budget}, which compares the task scores against the LWM baseline and a raw channel approach across different training budget percentages. We define the training budget as the specific fraction of the available training dataset used to fine-tune the downstream task heads; these subsets are sampled from the total allocated training splits (detailed in Table \ref{tab:dataset_partitioning}), while the validation and test sets remain completely fixed to ensure a fair and consistent evaluation benchmark. The comparative analysis reveals that the proposed model consistently outperforms both the LWM baseline and the raw channel approach across nearly all tasks and training budget levels. A significant finding is the model's high data efficiency; for instance, at a 25\% training budget, the proposed solution achieves a score of 51.50\% in beam prediction, surpassing the LWM baseline's performance at a much higher 50\% budget. This suggests that the latent representations learned during the MAE pretraining phase provide a robust foundation that requires fewer labeled samples to specialize for downstream applications. Notably, the LoS/NLoS classification task reports "N/A" results at the 10\% and 25\% levels. This is attributed to the extreme scarcity of data in this category with only 6 training samples available with further reduction by budget percentages results in insufficient data points (e.g., a single sample) to achieve stable model convergence or meaningful gradient updates. The results also highlight the performance of the raw channel approach, which utilized the proposed ResNet-based downstream heads directly on raw data, whereas the LWM baseline relied on standard linear layers. Interestingly, the raw approach demonstrated superior performance in the channel estimation task at lower budget levels (10\% to 50\%), suggesting that direct spatial features may be more readily exploitable by 2D CNN-based residual blocks in low-data regimes. However, as the data budget increases to 75\% and 100\%, the proposed model regains the peak estimation score of 99.118\%, indicating that the sophisticated feature extraction of the optimized encoder ultimately provides a more precise understanding of the wireless environment. 

\begin{figure}[t!]
\centering
\includegraphics[width=0.85\columnwidth]{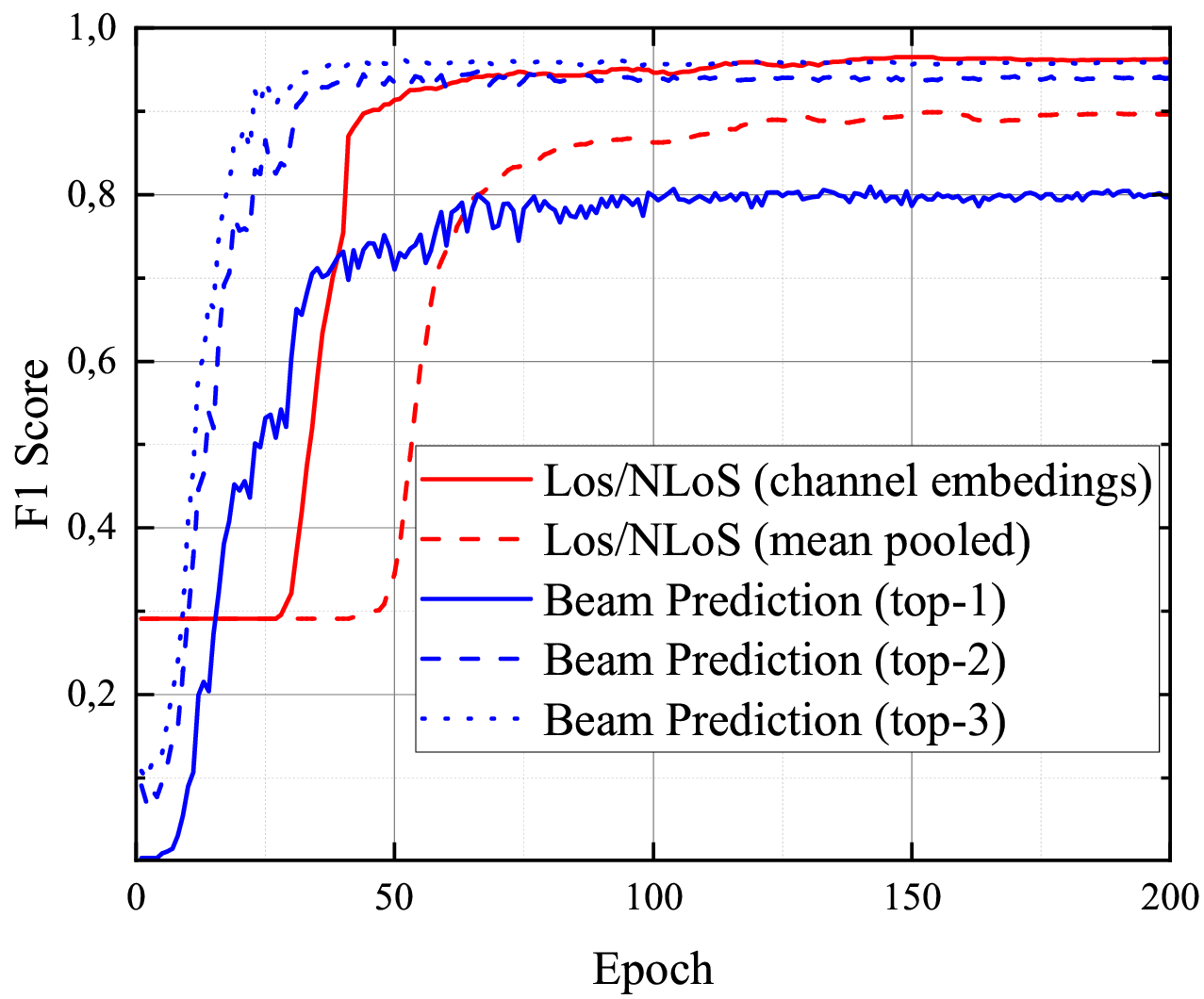}
\caption{Convergence behavior across LoS/NLoS classification and beam prediction tasks.}
\label{fig:losnlos_beamprediction}
\end{figure} 

\begin{figure}[t!]
\centering
\includegraphics[width=0.85\columnwidth]{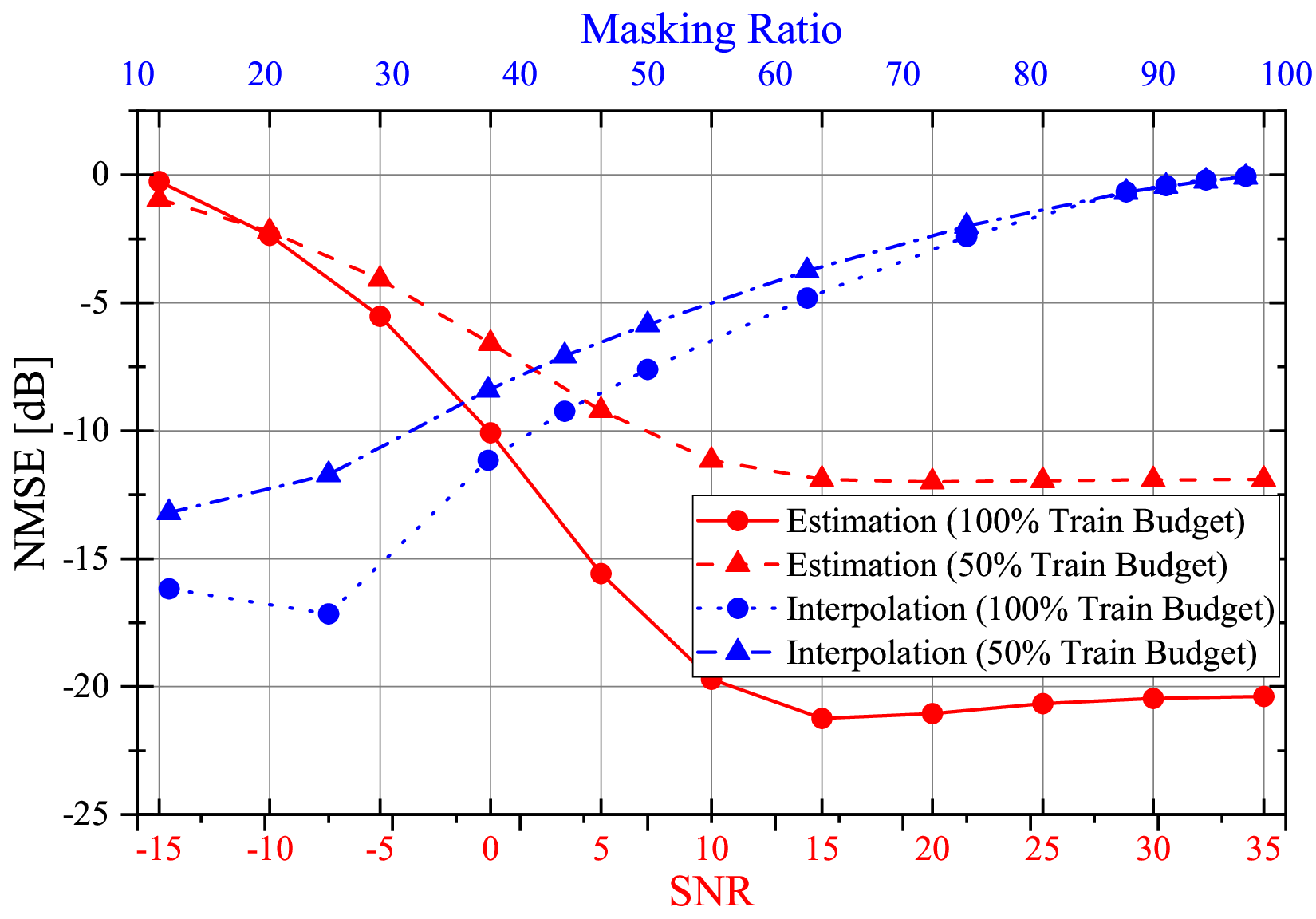}
\caption{Performance analysis across varying masking ratio and noise regimes for channel interpolation and channel estimation.}
\label{fig:interpolation_estimation}
\end{figure} 

\begin{figure}[ht!]
\centering
\includegraphics[width=0.95\columnwidth]{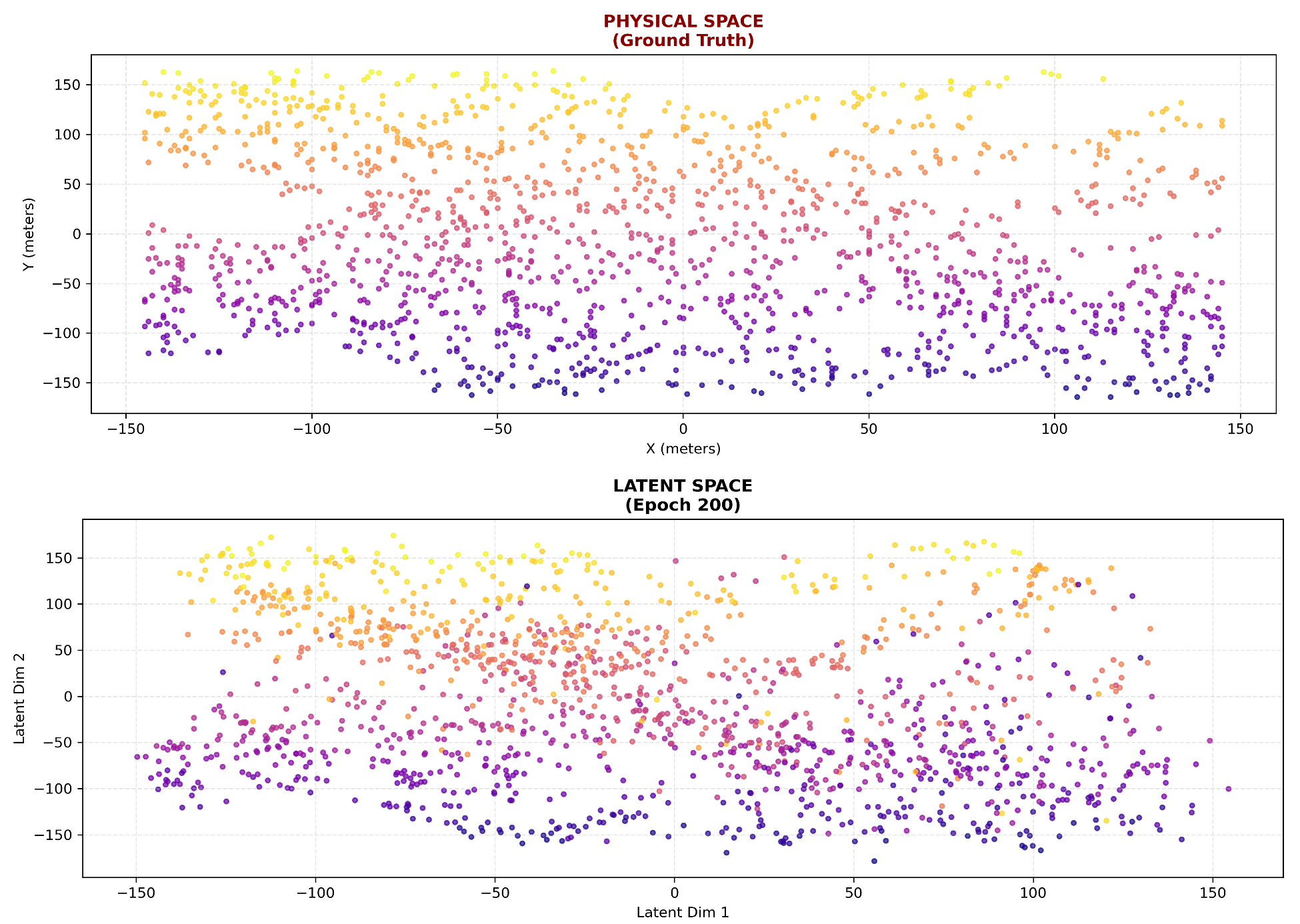}
\caption{Comparison of physical space coordinates and learned latent space representations in channel charting task.}
\label{fig:localization}
\end{figure} 

\begin{figure*}[ht!]
\centering
\includegraphics[width=2\columnwidth]{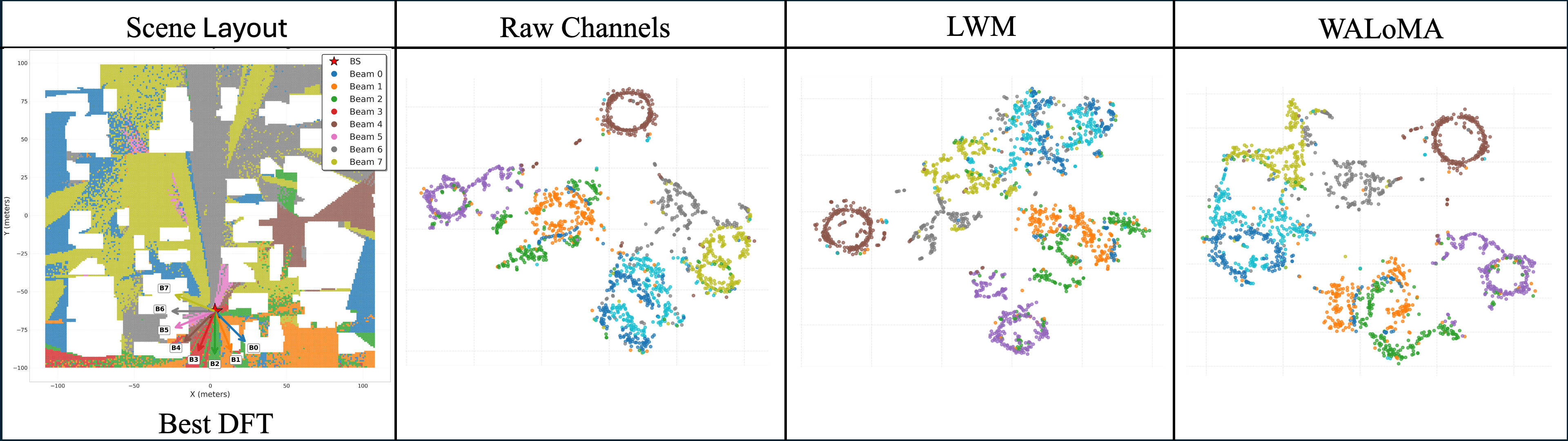}
\caption{Latent space t-SNE embeddings for the 28 GHz Miami scenario. On the left the physical scene layout with the optimal DFT beams. Sequentially, the panels evaluate the clustering performance by comparing the 2D representations of the raw channel data, the LWM baseline, and WALoMA.}
\label{fig:tsne}
\end{figure*} 

\subsection{Training Dynamics and Convergence Analysis}
The training curves illustrated in the Fig. \ref{fig:losnlos_beamprediction} provide a detailed view of the convergence behavior and stability of the proposed model across the beam prediction and LoS/NLoS classification tasks. For beam prediction, the model demonstrates rapid learning, reaching a peak Top-1 F1 score of 0.8045 within the first 100 epochs. As the prediction window expands, the performance increases significantly, achieving maximum F1 scores of 0.9473 for Top-2 and 0.9647 for Top-3 results. This trend indicates that while exact beam selection is a high-precision challenge, the model consistently identifies the optimal propagation paths within its top three candidates with very high reliability. In the LoS/NLoS classification task, the model reaches its peak performance of 0.9647 when utilizing the default channel embeddings. A notable observation is the performance disparity when switching to a mean-pooled token, which yields a significantly lower F1 score of 0.8985. The mean-pooled token is generated by averaging the latent embeddings across the entire sequence length, effectively compressing the high-dimensional spatial and frequency features into a single global vector. While this reduces the input dimensionality for the downstream head, the results suggest that this averaging process causes a loss of critical fine-grained information necessary for accurate environment classification. Furthermore, the fact that the mean-pooled approach underperforms relative to the baseline indicates that this specific feature extraction method may require a more specialized LoRA configuration, adjusting $r$ and $\alpha$ or transitioning to distinct downstream head architecture to effectively process the reduced information density of the pooled tokens. The superior results achieved using the full channel embeddings confirm that retaining the detailed spatial hierarchy learned by the encoder is essential for high-fidelity wireless sensing applications.

\subsection{Impact of Masking Ratio and SNR}
The performance of the proposed model in regression-based tasks is further analyzed in Fig. \ref{fig:interpolation_estimation}, which illustrates the NMSE for channel estimation and interpolation across different training budgets and environmental conditions. The channel interpolation task is evaluated against a patch-wise masking ratio, which represents the percentage of missing channel data the model must reconstruct. Utilizing the default 25\% masking ratio, the 100\% budget model achieves an NMSE of -17.16 dB, significantly outperforming the -11.71 dB achieved with a 50\% budget. An interesting phenomenon occurs at the 12.5\% masking ratio, where the 100\% budget model exhibits a slight worsening of NMSE (-16.17 dB) compared to its performance at 25\%. This localized dip is not present in the 50\% budget curve, suggesting a potential over-fitting to specific spatial structures or a sensitivity in the attention mechanism when the masking density is too low to engage the model's full reconstructive capabilities. As expected, performance for both budgets degrades as the masking ratio increases beyond 50\%, though the model continues to provide meaningful reconstructions even when more than 75\% of the channel patches are missing. The channel estimation results, plotted against varying SNR levels, demonstrate the high precision of the model in reconstructing channel state information. At the default SNR of 10 dB, the 100\% budget model achieves a significant NMSE of -19.71 dB, which continues to improve toward a near-perfect reconstruction floor of approximately -21 dB as SNR increases. Interestingly, the model maintains respectable performance even in lower SNR regimes, successfully capturing channel characteristics despite increased noise levels. A notable performance gap is observed when the training budget is reduced to 50\%; in this case, the NMSE plateaus around -12 dB for SNR levels above 15 dB. This indicates that while the encoder provides a strong feature base, the final fine-tuning of the residual heads for high-precision estimation is sensitive to the volume of task-specific training data.
\subsection{Latent Space Analysis and Topological Preservation}
The channel charting results in Fig. \ref{fig:localization} demonstrate that the learned latent space closely preserves the global topology and spatial consistency of the physical environment, as reflected by the alignment between the ground truth coordinates and the latent representations, including the consistent color-coded spatial gradients. Although minor distortions remain, the encoder effectively maps complex CSI into a structured manifold that captures the geographic distribution of users. Notably, under a multi-task setting and limited labeled data, the model achieves a charting score of 0.7718, significantly outperforming the baseline (0.6711), highlighting the effectiveness of the MAE-based framework with 2D positional encoding in learning spatially-aware and transferable channel representations.
\subsection{t-SNE Visualization of the Latent Space in mmWave Environments}
Fig. \ref{fig:tsne} illustrates the latent space representations for a Miami deployment scenario operating at 28 GHz, highlighting the robust cross-band generalization capabilities of the proposed architecture. To effectively visualize these high-dimensional embeddings, we employ t-distributed stochastic neighbor embedding (t-SNE). As a non-linear dimensionality reduction technique, t-SNE maps complex, multi-dimensional feature vectors into a 2D space by modeling the similarities between data points as joint probabilities. By minimizing the Kullback-Leibler (KL) divergence between the high-dimensional and low-dimensional probability distributions, t-SNE inherently preserves local data structures. Consequently, it forces similar channel features to cluster tightly together while pushing dissimilar, unrelated states further apart, making it highly effective for evaluating the discriminative power of a model's latent space. Applying this technique to the Miami scenario, the plots demonstrate highly effective transfer learning to \textbf{mmWave} environments, even though the foundation model was exclusively pre-trained on sub-6 GHz datasets but inferred on a 28 GHz dataset. For the beam prediction task ($N_B=8$), the proposed model exhibits superior spatial feature extraction compared to both the LWM baseline and the raw channel processing. The visualization reveals that the proposed framework visibly segregates each optimal discrete Fourier transform (DFT) beam class into distinct, tight clusters, whereas the baseline approaches produce significantly more scattered and overlapping manifolds. This well-structured latent space directly translates to predictive precision, allowing the model to achieve an 87\% top-1 accuracy for beam prediction. The clear class separation confirms that the model successfully captures fundamental geometric properties of the propagation channel and retains the detailed spatial hierarchies essential for high-fidelity wireless sensing.


 \section{Conclusion}
This work presented a robust foundation model framework tailored for diverse wireless communication tasks, leveraging an MAE architecture optimized with 2D PE and parameter-efficient fine-tuning via LoRA. By treating the wireless channel as a structured image-like grid, the model successfully learned high-fidelity latent representations that generalize across classification, regression, and spatial manifold learning tasks. The numerical results demonstrate that the proposed solution consistently outperforms established baselines, particularly in data-constrained regimes, while requiring significantly fewer trainable parameters. Despite the extreme scarcity of data in certain scenarios, such as LoS/NLoS classification, the model maintains high predictive accuracy and spatial awareness. Ultimately, this research underscores the potential of foundation models to serve as a versatile backbone for next-generation wireless systems, providing a scalable path toward unified and data-efficient network intelligence.

	\balance
	\bibliographystyle{ieeetr}
	\bibliography{mybibfile_final}
\end{document}